\documentclass[runningheads]{llncs}
\usepackage[T1]{fontenc}
\usepackage{graphicx}
\usepackage{amsmath}
\usepackage{amsfonts}
\usepackage{amssymb}
\usepackage{graphicx}
\usepackage{listings}
\usepackage{float}
\usepackage{caption}
\usepackage{subcaption}
\usepackage{hyperref}
\usepackage{color}
\usepackage{tablefootnote}
\usepackage{listings}
\usepackage{array}
\usepackage{fancyhdr}

\lstdefinestyle{mycodestyle}
{ 
    basicstyle=\small\ttfamily,
    numbers=left,
    stepnumber=1,
    showstringspaces=false,
    tabsize=1,
    breaklines=true,
    breakatwhitespace=false,
}

\title{Applying Formal Methods Tools to an Electronic Warfare Codebase (Experience report)}
\titlerunning{Applying Formal Method Tools}
\author{Letitia W. Li\inst{1}\orcidID{0000-0002-8373-6042} \and
Denley Lam\inst{1}\orcidID{0000-0002-4170-7147} \and
Vu Le\inst{1} \and
Daniel Mitchell \inst{1} \and
Mark J. Gerken \inst{1} \and
Robert B. Ross  \inst{1}
}

\institute{FAST Labs, BAE Systems, Arlington VA 22203, USA
\email{firstname.lastname@baesystems.us}}

\authorrunning{L. Li et al.}

\begin{document}

\maketitle

\begin{abstract}
While using formal methods offers advantages over unit testing, their steep learning curve can be daunting to developers and can be a major impediment to widespread adoption. To support integration into an industrial software engineering workflow, a tool must provide useful information and must be usable with relatively minimal user effort. In this paper, we discuss our experiences associated with identifying and applying formal methods tools on an electronic warfare (EW) system with stringent safety requirements and present perspectives on formal methods tools from EW software engineers who are proficient in development yet lack formal methods training. In addition to a difference in mindset between formal methods and unit testing approaches, some formal methods tools use terminology or annotations that differ from their target programming language, creating another barrier to adoption.   Input/output contracts, objects in memory affected by a function, and loop invariants can be difficult to grasp and use. In addition to usability, our findings include a comparison of vulnerabilities detected by different tools.  Finally, we present suggestions for improving formal methods usability including better documentation of capabilities, decreased manual effort, and improved handling of library code.

\end{abstract}

\section{Introduction}
\thispagestyle{fancy}
Formal methods is defined as the usage of mathematical models in system development \cite{10.1145/1592434.1592436}. Proofs offer rigorous evidence of system correctness across all possible execution paths, while unit testing can only show correctness over specified cases. Software flaws cost 30x more to fix post-production than if the flaw was detected in the design phase, where software patching cost an estimated 1.5 trillion dollars in the US in 2022 \cite{hackeroneCostSavings}. Studies have shown usage of formal methods ultimately decreases development cost of systems \cite{bowen2002ten}. However, formal methods is considered daunting for use by regular developers, especially with the high initial learning cost and even experts finding formal specifications difficult \cite{gore2013need,10.1145/3689374,broy2024does,kaleeswaran2023user}. In surveys of experts, the top barriers to adoption of formal methods in industry were “Engineers lack proper training in formal methods”, “Academic tools have limitations and are not professionally maintained”, “Formal methods are not properly integrated in the industrial design life cycle” and “Formal methods have a steep learning curve” \cite{garavel20202020}.  To this list we would add formal methods tools often have inconsistent/incomplete documentation and suffer from a lack of agreement on terminology between tools making it difficult to compare results across tools.

The Defense Advanced Research Projects Agency (DARPA) PROVERS program seeks to determine how to integrate formal methods tools into a conventional engineering development process for use by regular developers without intensive training \cite{darpaPROVERSPipelined}. Our role on the program is to 1) collect requirements and feedback from developers regarding different formal methods tools, and 2) determine how to integrate those tools into existing development workflow, in this case, an EW  system. From discussing with the engineering teams and sharing types of formal methods tools, we collected feedback on what kind of tools they would be willing to add to their workflow.

In this paper, we describe our experiences identifying formal methods tools for adoption into an engineering project with an existing codebase. We identified security needs, surveyed available tools, and identified the ones which addressed the security concerns we were interested in. Section \ref{sec:engineering_requirements} describes our system, the requirements collected from engineers, and a proposed summary of safety requirements in a taxonomy. Section \ref{sec:survey_of_open-source_c++_verification_tools} describes potentially applicable open-source formal methods tools. Section \ref{sec:experience_and_recommendations} describes the tools we selected and applied and their usage and capabilities, and recommendations for improvement. Section \ref{sec:related_work} describes the related work, and Section \ref{sec:conclusion} concludes the paper.

\section{Engineering Requirements}
\label{sec:engineering_requirements}

\subsection{System Overview}

The FAST Labs organization, which is the research and development branch of BAE System, actively maintains and develops a suite of electronic warfare applications \cite{militaryembeddedOpenArchitecture}. Each application is safety-critical, and must fulfill documented functionalities and be free of vulnerabilities. The codebase is in C++ with limited SysML models available. Currently, the code is run in a CI/CD pipeline implemented in Gitlab, with each commit checked against unit tests, and code scans by commercial software.

\subsection{System Requirements}

The documented requirements spanned a variety of domains, such as processing time, hardware resource usage, application interfaces, safety, and adherence to standards. To the best of our knowledge, modeling tools can estimate timing and resource usage, but not formally assure a function runs within a given time limit or with specified resource consumption. Many of these requirements are high-level of focus on functionality, and need to be further decomposed into specific security requirements to be addressed by formal methods tools. For example, the requirement that the code follows C++ best practices needs to be further decomposed into specific practices to be checked.

While certain system properties could be checked with model checkers, the lack of existing models for all system components was a barrier to rapid adoption of modeling tools. We considered generating SysML models for the selected codebase, but found it would require too much manual effort at this time. Tools implementing round-trip engineering, the ability to translate code to models and sync models with code updates, exist for certain specific cases and languages, but is yet to be broadly available \cite{marah2021model,10292497}. We found no tools available to translate C++ code to SysML models, so model checkers were of limited use to us.

\subsection{Safety properties of interest}

\begin{figure}[h!]
  \centering
  \includegraphics[width=0.9\textwidth]{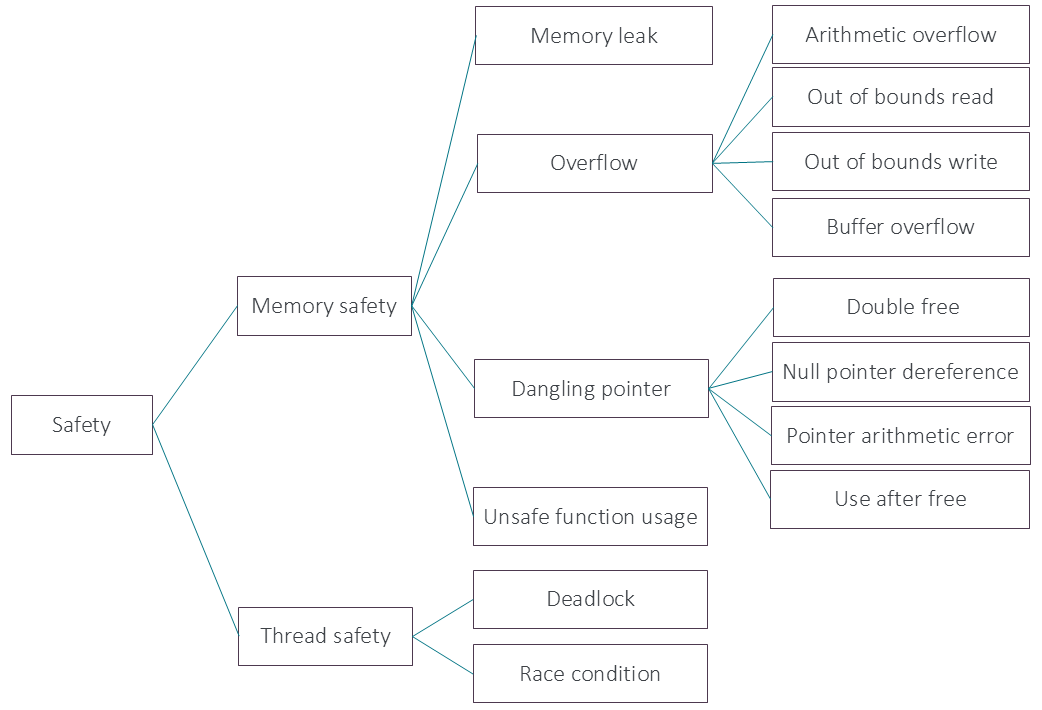}
  \caption{Vulnerability hierarchy}
  \label{fig:tree}
\end{figure}

While \cite{mitreMostDangerous} and \cite{codeintelligenceMostDangerous} offer lists of top security flaws, we found no existing taxonomy or list to cover all properties of interest in our system. Some common security flaws such as cross-site scripting, for example, did not apply in this case because the target app is not deployed online. Not all of the 699 Software Development CWEs were relevant to our system either, and we also wanted a more hierarchical and simplistic structure of multiple levels to display related errors together for easier readability for the engineer \cite{mitreCWE699Software}. Therefore, we summarized our software vulnerabilities of interest into a hierarchical tree as shown in Figure \ref{fig:tree} .

For the target EW app, the main C++ safety properties of interest include thread  and memory safety. For thread safety, we wish to avoid both deadlocks and race conditions~\cite{mitreCWE833Deadlock,mitreCWE362Concurrent}. As shown in Figure~\ref{fig:tree}, several memory safety issues were of concern:
\begin{itemize}
\item Memory leaks such as pointers not being deallocated after use; 
\item Address overflows including out of bounds reads or writes, buffer overflows, use of uninitialized variables, or other addressing problems leading to segmentation faults; 
\item Pointer problems including dereferencing null pointers, referencing out of scope pointers including use after free, double free errors, or general pointer arithmetic errors; and
\item Use of unsafe functions such as the insecure {\tt strcpy} function \cite{mitreCWE699Software}.
\end{itemize}

In addition to safety and memory issues, our engineers identified parsers specifically as a topic of interest. Parsers, handling untrusted external input, can be a major source of vulnerabilities, including the Heartbleed attack; recently (in 2025) vulnerabilities were identified in Go parsers and Apple Font parsers~\cite{bratus2017curing,trailofbitsUnexpectedSecurity,cyberpressFontParser}. Our engineering teams found parser debugging to be a challenging task. 
 They suggested they would be interested in tools which could verify parser implementations beyond just avoiding vulnerabilities such as buffer overflows. However, while multiple correct-by-construction parser generators exist \cite{blaudeau2020verified,diatchki2024daedalus,mundkur2020parsley,bangert2014nail,bratus2016implementing,delaware2019narcissus}, and we identified one parse result analyzer Polytracker \cite{sultanik2024polytracker}, as far as we could tell no tool exists that can determine if a parser implementation satisfies a given specification. 

\section{Survey of Open-source C++ Verification Tools}
\label{sec:survey_of_open-source_c++_verification_tools}

Our methodology involved searching google for ‘formal methods tool c++’ and ‘static analysis tools c++’, resulting in surveys of tools such as \cite{dwheelerHighAssurance} and individual tools. We attempted a literature search but decided not to include academic papers in our survey since few provided a link to their tool(s) and of those that did, many of them were dead. We focused on tools which could be applied to our codebase in C/C++ with a formal basis.  Instead, we focused on finding tools that had a formal basis and which could be applied to our C/C++ codebase. As the engineering team supporting development of the EW app had been clear that verification tools requiring them to learn new specification or annotation languages were of little interest, we excluded tools designed to operate on mathematical specifications such as the Prototype Verification System (PVS) and interactive theorem provers such as Coq/ACL2, all known to be challenging and time-consuming \cite{owre1999pvs,seo2022people,juhovsova2025pinpointing}.

We categorized the tools by criteria relevant for our choice of tool, similar to \cite{wikipediaListTools}, but narrowed the list to those that could support C++.  The resulting list is shown in Table~\ref{tab:tools}.  In addition  to the tool name and source, the table includes  additional details on usage and type of formal method(s) supported by that tool. The type of formal method used refers to the underlying mathematics or reasoning engine such as SAT solving, separation logic, Hoare Logic, or other approach. The type of interaction describes any required manual effort including annotations, compilation, or any other actions the user must take before using that tool. The last update date helps identify those tools that are actively being maintained and by extension would be more likely to support the most recent version of C++ and its libraries. Security class identifies the class of flaws addressed from the intermediate levels listed in the taxonomy provided earlier. Some tools, such as Kodiak, CN, and Frama-Clang allow the user to check user-defined properties, including if a variable falls within a specified range. 

\begin{table}[htbp]
    \centering
    \begin{tabular}{|m{1.5cm}|m{2.5cm}|m{2.5cm}|m{2cm}|m{2.7cm}|}
        \hline
        {\bf Tool}                 & {\bf Type of Formal Method} & {\bf Type of Interaction} & {\bf Last update\footnotemark[1] / Release} & {\bf Security Classes} \\
        \hline
        CBMC \cite{CBMC} & SAT solving & Command line interface with compilation & 07/09/2025 \newline 6.7.1 & Memory leak, \newline Overflow, \newline Dangling pointer \\
        \hline
        ESBMC \cite{ESBMC} & SAT solving & Command line interface & 10/12/2025 \newline 7.11 & Memory leak, \newline  Overflow,\newline   Dangling pointer, \newline Deadlock,\newline  Race condition \\
        \hline
        Clang Analyzer \cite{ClangAnalyzer} & Symbolic execution & Command line interface with compilation & 10/07/2025 \newline 21.1.3  & Memory leak, \newline  Overflow, \newline Dangling pointer,\newline  Unsafe function,\newline  Deadlock \\
        \hline
        CN \cite{CN} & Separation logic & Code annotation & 10/08/2025 \newline N/A & Memory leak, \newline  Overflow, \newline  Dangling pointer \\
        \hline
        Crux \cite{Crux} & Symbolic execution & Code annotation & 03/24/2025 \newline 0.10 & Overflow \\
        \hline
        Faial  \cite{faial} & SMT & Command line interface or Git action & 04/13/2024 \newline 1.0 & Race condition \\
        \hline
        Frama-Clang \cite{FramaClang} & Hoare logic (WP Plugin), Abstract interpretation (Eva Plugin) & Code annotation & 06/25/2025 \newline 0.0.18 & Memory leak,\newline  Overflow, \newline Dangling pointer,\newline  Deadlock, \newline Race condition \\
        \hline 
       IKOS \cite{IKOS} & Abstract interpretation & Command line interface & 12/31/2024 \newline 3.5 & Overflow, \newline Dangling pointer \\
        \hline
        Infer \cite{Infer} & Separation logic & Command line interface, Code annotation & 06/21/2024 \newline 1.2.0 & Memory leak, \newline  Overflow,\newline   Dangling pointer, \newline  Deadlock, \newline  Race condition \\
        \hline
        KLEE \cite{KLEE}  & Symbolic execution & Code annotation & 02/29/2024 \newline 3.1 & Overflow, \newline  Dangling pointer, \newline Deadlock, \newline Race condition \\
        \hline 
       Kodiak \cite{Kodiak} & Interval arithmetic, Bernstein expansion & C++ Library & 07/08/2022 \newline 2.0.4 & Overflow \\
        \hline 
       Seahorn \cite{Seahorn} & Horn clauses, Abstract interpretation & Code annotation, \newline  Command line interface & 09/12/2017 \newline 3.8 & Overflow, \newline Dangling pointer \\
        \hline
        SVF \cite{SVF} & Abstract execution & Command line interface on LLVM IR  & 06/05/2025 \newline 3.1 & Dangling pointer, \newline Overflow,  \newline Deadlock, \newline Race condition  \\
        \hline
    \end{tabular}
    \caption{Tools surveyed}
    \label{tab:tools}

\end{table}
\footnotetext[1]{As of October 18, 2025}

\section{Experience and Recommendations}
\label{sec:experience_and_recommendations}

In this section, we describe our experiences in selecting and applying formal methods tools on our system. From our survey and attempted integration of tools, we identified some recommendations for formal methods tool creators to expand their usability for more adoption by the general engineering community. From the list of tools, we assessed the ones we could easily install and run, and whose documentation gave us confidence that the tool would cover most of our safety requirements.  As a baseline non-formal tool, we used cppcheck, a free and open-source static analysis tool \cite{cppcheck}. Refining from the safety classes in Table \ref{tab:tools}, Table \ref{tab:safety_properties} shows each lowest-level safety property and if each of our listed tools handles each property.  As noted in the table, some properties of interest were only partially addressed by the given tool.

\begin{table}[htbp]
    \centering
    \begin{tabular}{|m{2.5cm}|m{1.6cm}|m{1.4cm}|c|c|c|c|c|}
        \hline
        {\bf Safety flaw} & {\bf Cppcheck} &   {\bf Frama-Clang} & {\bf CN} & {\bf Clang} &{\bf Infer} & {\bf IKOS} & {\bf ESMBC} \\
        \hline
        Memory leak \cite{mitreCWE401Missing}  & Y & Y & Y & Y & Y & N & Y \\
        \hline
        Arithmetic overflow \cite{mitreCWE190Integer} & Some &  Y & Some & Some & Some & Some & Some \\
        \hline
        Out of bounds read \cite{mitreCWE125Outofbounds} & Y & Y & Y & Y & Y & Y & Y \\
        \hline
        Out of bounds write \cite{mitreCWE787Outofbounds} & Y & Y & Y & Y & Y & Y & Y \\
        \hline
        Buffer overflow \cite{mitreCWE121Stackbased,mitreCWE122Heapbased} & Y & Y & Y & Y & Y & Y & Y \\
        \hline
        Double free errors \cite{mitreCWE415Double} & Y & Y & Y & Y & Y & Y & Y \\
        \hline
        Null pointer dereference \cite{mitreCWE476NULL} & Y & Y & Y & Y & Y & Y & Y \\
        \hline
        Pointer arithmetic error \cite{mitreCWE823Outofrange} & Y & Y & Y & Y & Y & Y & Y \\
        \hline
        Use after free \cite{mitreCWE416After} & Y & Y & Y & Y & Y & Y & Y \\
        \hline 
       Unsafe Function usage \cite{mitreCWE676Potentially}& Y  & N & N & Y & N & N & N \\
        \hline
        Deadlock \cite{mitreCWE833Deadlock}& N & Y & N & Some & Y  & N & Y \\
        \hline
        Race condition \cite{mitreCWE362Concurrent} & N   & Y & N & N & Y& N & Y \\
        \hline
    \end{tabular}
    \caption{Safety properties covered by tools}
    \label{tab:safety_properties}
\end{table}
\renewcommand{\thelstlisting}{\arabic{lstlisting}}

\subsection{Annotation}
Due in part to the challenging nature of the problem domain,   the software engineers supporting the target EW application expressed reluctance to learn new programming or annotation languages, but were interested to see if/how annotation-based methods could benefit their efforts. We therefore reviewed a few methods that relied on annotations so that we could get a sense of the relative complexity and level of effort required annotate source code and prove properties of interest using those tools. 

Annotation-based theorem provers, like Frama-Clang (the C++ version of Frama-C) and CN, verify software follows stringent safety checks as well as user-defined pre/post conditions. Given input to a function satisfies the pre-condition, the function must ensure the post-condition will always be true. These theorem provers check user-defined assertions and requirements instead of specific vulnerabilities alone.
For example, for a function $f(x)=z$ for sorting a sequence, we would specify a pre-condition that $x$ is an array which is allocated and initialized in memory, and a post-condition that the output $z$ is a permutation of the input $x$ (we can't drop or add values from the input sequence) and require $z$ to be ordered.  When using pre-/post-conditions, any implementation must satisfy the post-condition(s) whenever the input satisfies the input condition.  (Behavior is undefined in case the input does not satisfy the input condition.  Note that theorem provers may be able to show that an implementation satisfies its specification, but they may not necessarily show that the implementation terminates.  
In addition to pre- and post-conditions, annotation approaches generally support code assertions.  In the case of the sorting algorithm, we could add assertions concerning how the output $z$ is constructed.  An algorithm that randomly generates sequences then checks to see if the generated sequence is a sorted permutation of the input sequence would technically satisfy the post-condition.  Assertions could be used to require progress (incremental improvement in the solution), verify algebraic safety (over- or under-flows), or other properties of interest.  

Another barrier to adoption was the added time cost. Verifying the result of summing 0 to 9 inclusive in a simple loop in Frama-Clang of 5 lines required adding 7 lines of annotation, as shown in Listing \ref{framac}. The post-condition `ensures' requests that the verifier check that the returned value `result' is greater than 0. The other annotations are required by Frama-Clang: `assigns' tells the verifier which memory objects were modified by the function, the loop invariants indicate statements which are always true, the loop variant indicates variables or statements which change value during the loop, and loop assigns indicate which variables were modified during the loop \cite{furia2014loop}. One frustration we experienced was when Frama-Clang couldn’t make obvious conclusions without annotations. All the loop statements were necessary to enable the prover to verify the loop terminated, which should be obvious to even a novice programmer. Requiring engineers to provide significant numbers of annotations especially for relatively uninteresting portions of their algorithms was detrimental to future adoption and tended to decrease  confidence in annotation methods while increasing frustration in forcing manual annotation for a non-interesting verification results. CN uses similar annotaitons, but removes the need for loop variants and all `assigns' statements. In addition, CN requires additional stringent checks against overflow, such a check on multiplication to ensure the operation does not overflow, which other tools did not, as shown in Listing \ref{CN}.

\begin{lstlisting}[caption=Frama-C annotations to verify loop, label={framac}]

/*@
        ensures \result > 0;
        assigns \nothing;
*/
int loop(){
        int j=0;
        /*@
        loop invariant 0 <= i <=10;
        loop invariant 0<=j<=i*10;
        loop invariant (i-1) <=j;
        loop assigns i,j;
        loop variant (10-i);
*/
        for (int i=0; i<10; i++){
                j+=i;
        }
        return j;
} 
}
\end{lstlisting}

\begin{lstlisting}[caption={CN checks to avoid multiplication overflow}, language=C++,label={CN}]
int test(int a, int b){
        if (a==0 || b==0){
                return 0;
        }
        else if (a > 0 && b>0 && INT_MAX / b > a){
                return a*b;
        }
        else if (a>0 && b <0 && b > INT_MIN/a){
                return a*b;
        }
        else if (a<0 && b>0 && a > INT_MIN / b){
                return a*b;
        }
        else if (a<0 && b<0 &&  b > INT_MAX / a){
                return a*b;
        }
        return 0;
}
\end{lstlisting}
While they allow the developer to verify custom requirements, the pre-/post-condition format of these annotations require a different method of thinking than engineers are used to, having to translate unit tests into comprehensive statements covering all cases. In fact, we also had no documented requirements for individual functions. 
Another barrier to adoption was the added time cost. Verifying a simple loop in Frama-Clang of 5 lines required adding 7 lines of annotation, as shown in Listing \ref{framac}. One frustration we experienced was when Frama-Clang couldn’t make obvious conclusions without annotations. All the loop invariants were necessary to enable the prover to verify the loop terminated, which should be obvious to any even novice programmer, decreasing confidence about the prover and increasing frustration in forcing manual annotation for a non-interesting verification result.

\subsection{Static Analysis}
\label{sec:static}
Static analysis tools required lower manual effort while still detecting flaws in our codebase, as they are run on the code as is without requiring any manual annotations.  However, as shown in Table~\ref{tab:safety_properties} no single tool addressed all vulnerabilities of interest; each focused on a different set of vulnerabilities. For example, integer overflow was covered by most tools, but floating point overflow was only covered by Frama-Clang. Only Frama-Clang, IKOS, and ESBMC could cover the multi-threading related properties of catching deadlocks and race conditions. The pointer-related vulnerabilities, on the other hand, were handled by all our evaluated tools.

Unfortunately, a lack of consistent interpretation of results makes it difficult to compare tools. For example, is a double free (something that can lead to undefined behavior) an extension of a use after free? Some tools consider them similar bug classes. Until recently even the output formats are custom. Tools are adopting Static Analysis Results Interchange Format (SARIF) as a common output format, which makes comparing results significantly easier \cite{sarif}. SARIF, supports three severities; error, warning, and note. Identifying critical bugs should become easier with a common language.

\begin{figure}[h!]
  \centering
  \includegraphics[width=0.95\textwidth]{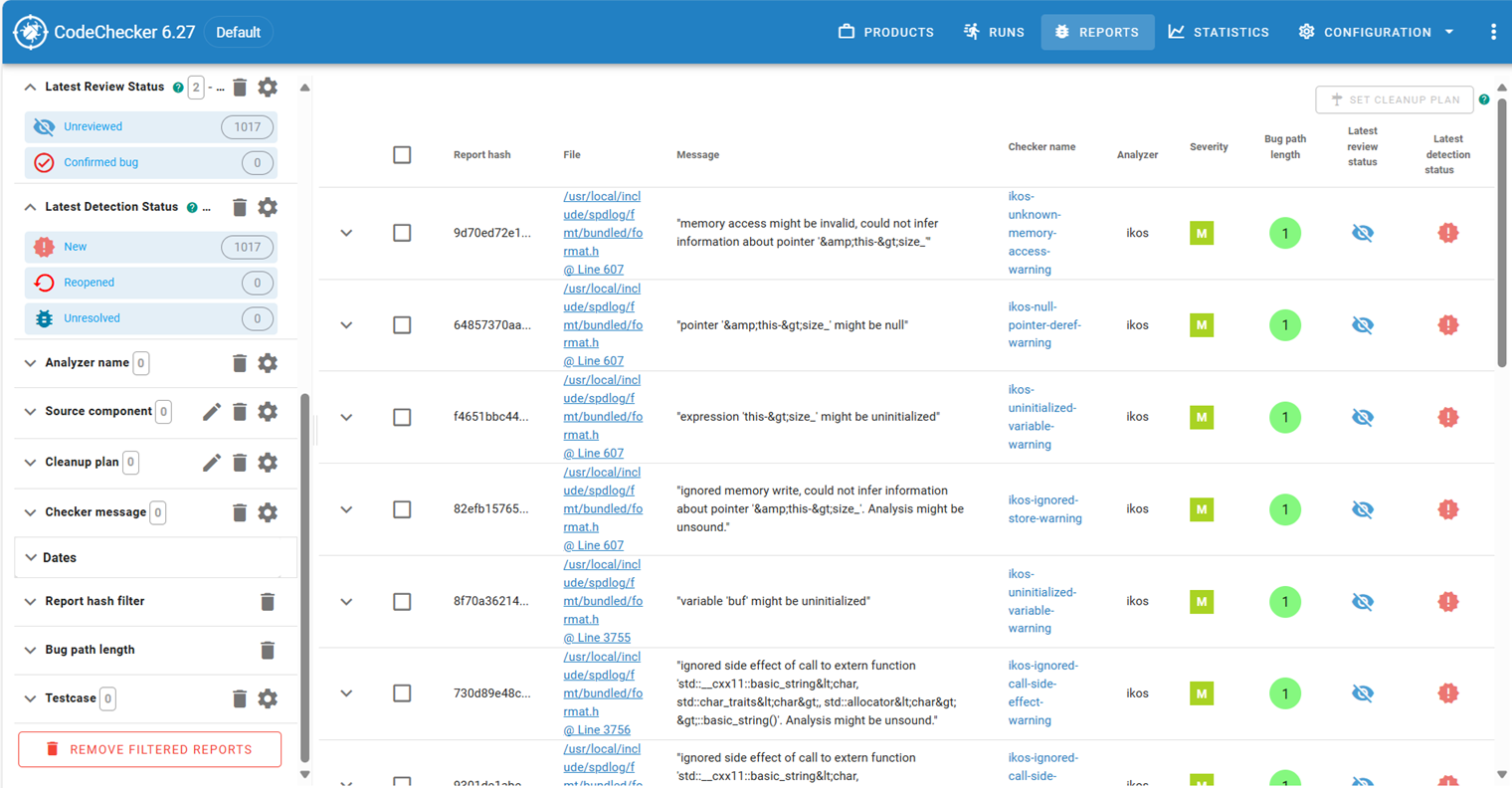}
  \caption{CodeChecker UI}
  \label{fig:codechecker}
\end{figure}

For combining the results, we used CodeChecker, an infrastructure for managing results from multiple static analysis tools \cite{CodeChecker}. In total, we had over 160k detections, with the vast majority originating from IKOS, and ~40k detections originating from library code. Figure \ref{fig:codechecker} shows a screenshot of the online report interface, for the spdlog logging library file format.h. Note that several reported vulnerabilities include wording that the reported vulnerability {\it might} exist, such as ``variable `buf' might be unitialized.''  In these cases, a manual code review would be needed to verify those findings.

We demonstrate the difference on a base 64 decode file shown in Listing \ref{b64} with multiple known bugs: potential reads and writes out of bounds of arrays in b64\_decode, a possible buffer overflow in line 52 due to the unsafe strcpy function if argv[0] is longer than 20 characters, and a memory leak in not freeing `frame' at the end of main. Cppcheck and infer both found no bugs in the code, while the clang-tidy analyzer returned 3 warnings as shown in Listing \ref{b64_clang}, ESBMC returned 1 warning as shown in Listing \ref{b64_esbmc}, and IKOS returned 17 warnings with an excerpt shown in Listing \ref{b64_ikos}. Despite all of these tools claiming to detect buffer overflow and out of bounds errors, we still received different analysis results from each, with IKOS being the most rigorous though it did not detect the memory leak that Clang-tidy did. CN similarly required bounds checks on the array before accesses, but also required checks that no mathematical operations (such as in[i]-43) would overflow or underflow. Even for a well-known vulnerability from the insecure strcpy function, it is interesting to note that analyzers did not agree on vulnerabilities detected.

\subsection{CI/CD Integration}
An important useability trait is how well the tool integrates into the CI/CD pipeline and on performance. For example, cppcheck analyzes on a file level basis and can be provided a directory to check all files within. IKOS needs to be integrated into the build infrastructure because it uses clang to generate intermediate bitcode for later analysis. Another method used by Infer is to operate on the compile\_commands.json, generated post-build by the compiler. Typically, deeper integration is slower and requires more work, but hopefully with the benefit of better results. Through the lens of a developer any method is valuable if the performance loss is offset by gains in verification. The time gap can be significant, ranging from a matter of minutes to hours. Given this performance loss in formal method tools, there is a higher burden to provide verification that is useful, actionable, and free of significant false positives.

\begin{lstlisting}[caption=Frama-C annotations to verify loop, label={b64}, style=mycodestyle]
#include <cstddef>
#include <cstring>

int b64invs[] = { 62, -1, -1, -1, 63, 52, 53, 54, 55, 56, 57, 58,
 ...

int b64_isvalidchar(char c) {
        if (c >= '0' && c <= '9')
                return 1;
...
}

int b64_decode(const char *in, unsigned char *out) {
        size_t len, i, j;
        int v;
        if (in == NULL || out == NULL)
                return 0;
        len = strlen(in);
        for (i=0; i<len; i++) {
                if (!b64_isvalidchar(in[i])) {
                        return 0;
                }
        }
        for (i=0, j=0; i<len; i+=4, j+=3)  {
                v = b64invs[in[i]-43];
                v = (v << 6) | b64invs[in[i+1]-43];
                v = in[i+2]=='=' ? v << 6 : (v << 6) | b64invs[in[i+2]-43];
                v = in[i+3]=='=' ? v << 6 : (v << 6) | b64invs[in[i+3]-43];
                out[j] = (v >> 16) & 0xFF;
                if (in[i+2] != '=')
                        out[j+1] = (v >> 8) & 0xFF;
                if (in[i+3] != '=')
                        out[j+2] = v & 0xFF;
        }
        return 1;
}

int main(int argc, char *argv[]) {
        char* frame=new char[20];
        unsigned char* output= new unsigned char[20];
        strcpy(frame, argv[0]);
        b64_decode(frame, output);
}
\end{lstlisting}

\begin{lstlisting}[caption=Clang Analyzer analysis of base 64 decode, label={b64_clang}]
3 warnings generated.
b64_decode.cpp:41:1: warning: Call to function `strcpy' is insecure as it does not provide bounding of the memory buffer. Replace unbounded copy functions with analogous functions that support length arguments such as `strlcpy'. CWE-119 [clang-analyzer-security.insecureAPI.strcpy]
strcpy(frame, argv[0]);
^
b64_decode.cpp:43:1: warning: Potential leak of memory pointed to by `frame' [clang-analyzer-cplusplus.NewDeleteLeaks]
}
^
b64_decode.cpp:39:13: note: Memory is allocated
char* frame=new char[20];
            ^
b64_decode.cpp:43:1: note: Potential leak of memory pointed to by `frame'
}
^
b64_decode.cpp:43:1: warning: Potential leak of memory pointed to by `output' [clang-analyzer-cplusplus.NewDeleteLeaks]
}
^
b64_decode.cpp:40:24: note: Memory is allocated
unsigned char* output= new unsigned char[20];
                       ^
b64_decode.cpp:43:1: note: Potential leak of memory pointed to by `output'
}
^
\end{lstlisting}

\begin{lstlisting}[caption=ESBMC analysis of base 64 decode, label={b64_esbmc}]
State 6 file string.c line 30 column 3 function strcpy thread 0
----------------------------------------------------
Violated property:
  file string.c line 30 column 3 function strcpy
  Source pointer is null
  src != (signed char *)0
\end{lstlisting}

\begin{lstlisting}[caption=IKOS analysis of base 64 decode, label={b64_ikos}]
b64_rev.cpp: In function `b64_decode(char const*, unsigned char*, unsigned long)':
b64_rev.cpp:25:21: warning: possible buffer overflow, could not bound index for access of global variable `b64invs' of 80 elements
                v = b64invs[in[i]-43];
                    ^
b64_rev.cpp: In function `b64_decode(char const*, unsigned char*, unsigned long)':
b64_rev.cpp:26:32: warning: possible buffer overflow, pointer `&b64invs[(int64_t)(((int32_t)in[i + 1]) - 43)]' accesses 4 bytes of global variable `b64invs' of size 320 bytes
                v = (v << 6) | b64invs[in[i+1]-43];
                               ^
...
b64_rev.cpp: In function `b64_decode(char const*, unsigned char*, unsigned long)':
b64_rev.cpp:33:34: warning: possible buffer overflow, pointer `&out[j + 2]' accesses 1 bytes of dynamic memory allocated at `main:38:24' of size 20 bytes
                        out[j+2] = v & 0xFF;
                                 ^
b64_rev.cpp: In function `main':
b64_rev.cpp:41:1: warning: expression `*argv' might be uninitialized
strcpy(frame, argv[0]);
^
b64_rev.cpp: In function `main':
b64_rev.cpp:41:1: warning: pointer `*argv' might be null
strcpy(frame, argv[0]);
^
b64_rev.cpp: In function `main':
b64_rev.cpp:41:1: warning: memory access might be invalid, could not infer information about pointer `*argv'
strcpy(frame, argv[0]);
^
b64_rev.cpp: In function `main':
b64_rev.cpp:41:15: warning: possible buffer overflow, pointer `argv' accesses 8 bytes at offset 0 bytes of `argv'
strcpy(frame, argv[0]);
              ^
\end{lstlisting}
\subsection{Feedback and Recommendations}
The main engineering requirement and feedback was to focus on making the tools as easy to use as possible. From the engineer’s perspective, they are open to adopting new tools only with a clear benefit in terms of time and cost. Learning a new language (e.g., tool-specific code annotations) was considered too much of a time commitment. There were also concerns about  the time it takes to annotate and debug annotations; across a large codebase annotations were considered too time-consuming.  The tools mainly of interest to engineers were therefore static analysis tools with no manual effort on the part of the user, especially those capable of detecting high interest software flaws. However, annotation-based tools enforced a rigor of avoiding undefined behavior, so with some modification or automation, we believe they eventually could be more generally applicable. We next suggest improvements to formal methods to become more user-friendly to developers.

\subsubsection{Engineering Terminology}
Engineers had difficulty with the formal language concepts not present in common programming languages. The annotation language, like ACSL, was unfamiliar to engineers, and contained keywords that don’t exist in common programming languages. For example, our engineers were confused about ACSL’s \textbackslash assigns, \textbackslash valid, and \textbackslash old, all constructs that are not present in common programming languages. We understand formal languages require additional constructs than regular programming language, but we suggest re-using existing terminology when possible, such as [0, n] for array ranging instead of CN’s ‘array\_shift’. 
In addition, behavioral contracts in terms of pre- and post-conditions used in tools like Frama-Clang are a novel way of reasoning about programs that don't necessarily align with the way engineers think about programs; they tend to think in terms of test cases, not in terms of memory objects a function may affect. The level of requirements themselves can be an impediment to formal methods adoption.  For the EW application used in this study, we did not have any requirements which could be translated directly into ACSL contracts, requiring us to manually determine them. Instead, requirements were expressed at a system level, not by function. (Requirements at the individual function level were defined inside the issue tracker integrated into the overall DevOps process.)  While an ensemble of unit tests can be summarized into input/output contracts with effort, requirements generally don’t have any summary on memory objects. Beyond using annotation languages closer to common programming languages, engineers could benefit from automation of annotation generation. With the rise of AI assistants and a certain predictability in annotations, it would be convenient for the tool to automatically generate annotations including loop invariants when feasible. Some level of annotation automation would make annotation tools more accessible.

\subsubsection{Improved Documentation of capabilities}
Completing the information contained in  Table \ref{tab:tools} took significant manual effort, especially concerning identifying the security classes covered by particular tools; this information was not always listed  on their description page. Given the number of available tools, it is unrealistic for an engineering team to download and try each available formal methods tool. Even if there exist unevaluated tools who could cover all our safety requirements, we were less likely to try them due to the difficulty in finding their capabilities easily. Infer, for example, helpfully documented all their vulnerabilities covered with examples, though that information was not separated by programming language. The examples helped as terminology for certain vulnerabilities can vary. Even more convenient for the user, however, would be to list the CVEs covered by the tool explicitly. ESBMC used terminology most closely matching CVEs in their documentation.
While we suggested a software safety taxonomy, each tool uses their own terminology. Terminology standardization would simplify tool  comparison and help identify which combination of tools covers  vulnerability classes of interest. 

\subsubsection{Handling library code}
As we showed in Section \ref{sec:static}, warnings and errors in library code were reported in analysis of compiled binaries.  It was difficult to sort out which were errors in our source code that could be addressed and which belonged to library code. At the same time, annotation-based theorem provers could not handle code including common libraries, including {\tt sys/time.h}, and associated common functions such as {\tt std::min}. We understand that annotating and verifying all libraries would be extremely time consuming, but at the same time a function using those library functions cannot be verified unless the library functions are proven correct. Instead, we suggest allowing tools to ignore library functions and assume them correct, and indicating these assumptions. 

In CodeChecker, nearly a fourth of reported errors resided in library headers. The results can be filtered out, but they unnecessarily inflate the number of detections.  Additionally, some of the reports were false reports of unreachable code by IKOS, such as on the C++ Utilities libraries. 

From our survey, we further evaluated seven static analysis tools, and discussed our experiences with the type types: annotation-based theorem proving and static analysis. We were able to cover all of our vulnerability classes with a combination of tools, and we suggest a common standardized vocabulary would allow us to better compare tools. A few usability improvements would help these tools better integrate into an engineering workflow, including filtering out library code, easier annotation languages, and automated annotation systems.    

\section{Related Work}
\label{sec:related_work}

In our survey of existing works, we found compilations of formal methods tools, recommendations for increasing adoption of formal methods tools, and usability studies of tools, but no papers detailing a combined survey and application of formal methods tools with recommendations for industrial adoption. 
\cite{dwheelerHighAssurance} provides an extensive list of open-source formal methods tools, with some overlap with our list, while also including theorem provers and proof checkers. \cite{kulik2022survey} provides an extensive overview of formal methods tools used for a variety of domains but focused only on security. \cite{almeida2011overview} provides a comprehensive overview of formal methods techniques and associated tools, with some examples of usage and \cite{foster2023teaching}  identifies a list of formal methods tools by category, but only surveys tools taught in courses. \cite{arusoaie2017comparison} not only surveyed tools detecting vulnerabilities in C++ code and evaluated them against the Toyota ITC test suite benchmark calculating false and true positive rate. \cite{black2020sate} similarly performed a comparison between ASTREE and Frama-C. While we have listed some of the same tools such as Frama-Clang, Clang, and Infer, we focused on open-source formal methods tools only and user experience rather than vulnerability detection alone.
 
\cite{garavel20202020} surveys expert users of formal methods. The survey reports most experts similarly recommend improving tool usability and comprehensibility. Our paper, however, focuses on the perspective of engineers instead of formal methods experts. \cite{gleirscher2023manifesto} provides suggestions for improving formal methods adoption in industry, with similar suggestions to ours about making tools more usable and automated. These papers, however, focus on an overview of the field of formal methods and summary of success stories on formal methods while we focus only on our experiences. 

\cite{10207560} details formal methods usage specifically in the aerospace domain. \cite{ter2024formal} also provides a comprehensive survey of usage of formal methods in industry, with different tools used by each domain, including ours of aerospace. Our paper, however, focuses specifically on C++ tools most accessible to engineers applicable to a more specific electronic warfare application. \cite{kaleeswaran2023user} provides an experience report of use of model checkers of automotive engineers in Bosch. Their work involves a formal survey rating difficulty, time spent, etc while our work focuses on feedback from a single team without an extensive user study.
Our paper confirms and builds on these suggestions for increased usability of formal methods by developers, with additional perspective of practical needs by engineers.

\section{Conclusion}
\label{sec:conclusion}

Formal methods tools provide powerful analysis capabilities, but many are inaccessible or impractical for standard software engineers. This paper provides a case study of the engineer’s perspective in identifying and applying formal methods tools on an existing C++ codebase of an electronic warfare system with stringent safety requirements. We present a survey of applicable tools and characterize them on the vulnerabilities each address and manual steps required to run the tool. For improved adoption of formal methods in engineering projects, we suggest clear documentation on tool capabilities, decrease in manual effort, and improved isolation of library code. We also note a need for parser analysis tools and code to model translators, especially for generating C++.

Future work on this program involves more detailed investigation into differences in detection of vulnerabilities, such as using the Toyota ITC benchmark, along with identification of other formal methods tools usable by engineers. With the recent development of automated assistance for interactive theorem proving, we will evaluate the more developed assistants to identify if any are understandable \cite{10.1145/3691620.3695521,10.1145/3691620.3695357} . This assessment will better help engineering teams determine which formal methods tools can best support improved safety of their systems without major manual effort.

\begin{credits}
\subsubsection{\ackname} The authors would like to thank Dr. Ed Hill for his perspective and feedback in support of this paper. This material is based upon work supported by the Air Force Research Laboratory and DARPA under Contract No. FA8750-24-C-B035. Any opinions, findings and conclusions or recommendations expressed in this material are those of the authors and do not necessarily reflect the views of the Air Force Research Laboratory and DARPA.
\end{credits}

\bibliographystyle{splncs04}
\bibliography{fm_paper}

\end{document}